\documentclass[prd,preprintnumbers]{revtex4}

\usepackage{amsmath}
\usepackage{amsfonts}
\usepackage{graphicx}

\begin{document}
\tolerance=5000
\def\pp{{\, \mid \hskip -1.5mm =}}
\def\cL{{\cal L}}
\def\be{\begin{equation}}
\def\ee{\end{equation}}
\def\bea{\begin{eqnarray}}
\def\eea{\end{eqnarray}}
\def\tr{{\rm tr}\, }
\def\nn{\nonumber \\}
\def\e{{\rm e}}

\preprint{}

\title{Dark energy and possible alternatives}

\author{ M. Sami}
\affiliation{Center of Theoretical Physics, Jamia Millia Islamia,
Jamia Nagar, Delhi-110092, India}

\date{{\small \today}}


\begin{abstract}
We present a brief review of various approaches to late time
acceleration of universe. The cosmological relevance of scaling
solutions is emphasized in case of scalar field models of dark
energy. The underlying features of a variety of scalar field models
is highlighted. Various alternatives to dark energy are discussed
including the string curvature corrections to Einstein-Hilbert
action, higher dimensional effects, non-locally corrected gravity
and $f(R)$ theories of gravity. The recent developments related to
$f(R)$ models with disappearing cosmological constant are reviewed.
\end{abstract}


\maketitle


\section{Introduction}
The accelerated expansion has played a very important role in the
history of our universe. Universe is believed to have passed through
inflationary phase at early epochs and there is a growing faith that
it is accelerating at present. The late time acceleration of the
universe, which is directly supported by supernovae observations,
and indirectly, through observations of the microwave background,
large scale structure and its dynamics, weak lensing and baryon
oscillations, poses one of the most important challenges to modern
cosmology.

Einstein equations in their original form, with an energy-momentum
tensor for standard matter on the right hand side, cannot account
for the observed accelerated expansion of universe.  The standard
lore aimed at capturing this important effect is related to the
introduction of the energy-momentum tensor of an exotic matter with
large negative pressure dubbed {\it dark energy} in the Einstein
equations. The simplest known example of dark energy (for recent
reviews, see \cite{review}) is provided by the cosmological constant
$\Lambda$. It does not require {\it adhoc} assumption for its
introduction, as is automatically present in the Einstein equations,
by virtue of the Bianchi identities.

The field theoretic understanding of $\Lambda$ is far from being
satisfactory. Efforts have recently been made to obtain $\Lambda$ in
the framework of string theory, what leads to a complicated
landscape of de-Sitter vacua. It is hard to believe that we happen
to live in one of the $10^{500}$ vacua predicted by the theory. One
might take the simplified view that, like $G$, the cosmological
constant $\Lambda$ is a fundamental constant of the classical
general theory of relativity and that it should be determined from
large scale observations. It is interesting to remark that the
$\Lambda CDM$ model is consistent with observations at present.
Unfortunately, the non-evolving nature of $\Lambda$ and its small
numerical value lead to a non-acceptable fine-tuning problem. We do
not know how the present scale of the cosmological constant is
related to Planck's or the supersymmetry breaking scale; perhaps,
some deep physics is at play here that escapes our present
understanding.

The fine-tuning problem, associated with $\Lambda$, can be
alleviated in scalar field models which do not disturb the thermal
history of the universe and can successfully mimic $\Lambda$ at late
times. A variety of scalar fields have been investigated to this
end\cite{review}; some of them are motivated by field/string theory
and the others are introduced owing to phenomenological
considerations. It is quite disappointing that a scalar field
description lacks predictive power; given {\it a priori} a cosmic
evolution, one can always construct a field potential that would
give rise to it. These models should, however, not be written off,
and should be judged by the generic features which might arise from
them. For instance, the tracker models have remarkable features
allowing them to alleviate the fine-tuning and coincidence problems.
Present data are insufficient in order to conclude whether or not
the dark energy has dynamics; thus, the quest for the metamorphosis
of dark energy continues\cite{bruce}

One can question the standard lore on fundamental grounds. We know
that gravity is modified at small distance scales; it is quite
possible that it is modified at large scales too where it has never
been confronted with observations directly. It is therefore
perfectly legitimate to investigate the possibility of late time
acceleration due to modification of Einstein-Hilbert action. It is
tempting to study the string curvature corrections to Einstein
gravity amongst which the Gauss-Bonnet correction enjoys special
status. A large number of papers are devoted to the cosmological
implications of string curvature corrected
gravity\cite{fr,NOS,KM06,TS,CTS,Neupane,Cal,Sami06,Annalen,Sanyal}.
These models, however, suffer from several problems. Most of these
models do not include tracker like solution and those which do are
heavily constrained by the thermal history of universe. For
instance, the Gauss-Bonnet gravity with dynamical dilaton might
cause transition from matter scaling regime to late time
acceleration allowing to alleviate the fine tuning and coincidence
problems. But it is difficult to reconcile this model with
nucleosynthesis\cite{TS,KM06}constraint. The large scale
modification may also arise in extra dimensional theories like DGP
model which contains self accelerating brane. Apart from the
theoretical problems, this model is heavily constrained by
observation.

On purely phenomenological grounds, one could seek a modification of
Einstein gravity by replacing the Ricci scalar by generic function
$f(R)$\cite{review1,FRB,FRB1}. The $f(R)$ gravity theories giving
rise to cosmological constant in low curvature regime are faced with
difficulties which can be circumvented  in $f(R)$ gravity models
proposed by Hu-Sawicki and Starobinsky \cite{HS,star} (see
Ref.\cite{App1} on the similar theme). These models can evade solar
physics constraints by invoking the chameleon mechanism
\cite{HS,star,Tsujikawa}. An important observation has recently been
made in Refs.\cite{App2,frolov}(see also Ref.\cite{others} on the
related theme), namely, the minimum of scalaron potential which
corresponds to dark energy can be very near to $\phi=0$ or
equivalently $R=\infty$. As pointed out in Ref.\cite{Tsujikawa}, the
minimum should be near the origin for solar constraints to be
evaded. Hence, it is most likely that we hit the singularity if the
parameters are not properly fine tuned. This may have serious
implications for relativistic stars\cite{maed1}.\\ In what follows
we shall briefly review the aforesaid developments.

\section{Late time acceleration and cosmological constant}
 Einstein equations exhibit simple analytical solutions in a
 homogeneous and isotropic universe. The dynamics in this case
 is described by a single function of time $a(t)$ dubbed {\it scale
 factor},
\begin{eqnarray}
&& H^2 \equiv \frac {\dot{a}^2}{a^2}=\frac {8\pi G \rho}{3}-\frac {K}{a^2} \, \nonumber \\
\label{Hubbleeq}
 &&\dot{\rho}+3 H(\rho+p)=0\,,\nonumber
\end{eqnarray}
where $\rho$ is designates the total energy density in the universe.
Three different possibilities, $K=0$, $K >0$ or
 $K<0$
 correspond to flat geometry, hyperbolic geometry and geometry of
 the sphere correspondingly. Evolution can not change the nature of
 a particular geometry. What geometry we live in, depends upon the
 energy content of the universe,
\begin{eqnarray}
&&\frac{K}{a^2}=H^2 \left(\Omega(t)-1\right) \nonumber \\
&&\Omega=\rho/\rho_c,~~~\rho_c=3H^2/8 \pi G \nonumber
\end{eqnarray}
Observations on CMB reveal that we live in a nearly critical
universe, $K=0$ or $\rho=\rho_c$ which is consistent with
inflationary paradigm. The equation for acceleration has the
following form,
\begin{eqnarray}
&&\frac{\ddot{a}}{a}=-\frac{4 \pi G}{3} \left(\rho +3p\right) \nonumber\\
&&  \ddot{a} > 0 \Longleftrightarrow p < -\frac{\rho}{3}:~ Dark
Energy.\nonumber
\end{eqnarray}
Thus an exotic fluid with large negative pressure is needed to fuel
the accelerated expansion of universe. Let us note that pressure
corrects the energy density and positive pressure adds to
deceleration where as the negative pressure contributes towards
acceleration. It might look completely opposite to our intuition
that highly compressed substance explodes out with tremendous impact
whereas in our case pressure acts in the opposite direction. It is
important to understand that our day today intuition with pressure
is related to pressure force or pressure gradient. In a homogeneous
universe pressure gradients can not exist. Pressure is a
relativistic effect and can only be understood within the frame work
of general theory of relativity. Pressure gradient might appear in
Newtonian frame work in an inhomogeneous universe but pressure in
FRW background can only be induced by relativistic effects. Strictly
speaking, it should not appear in Newtonian gravity; its
contribution is negligible in the non-relativistic limit. Indeed, in
Newtonian cosmology, acceleration of a particle on the surface of a
homogeneous sphere with density $\rho$ and radius $R$ is given by,
\begin{equation}
\ddot{R}=-\frac{4\pi}{3}G\rho R
\end{equation}
The simplest possibility of dark energy is provided by cosmological
constant which does not require an {\it adhoc} assumption for its
introduction; it is automatically present in Einstein equations by
virtue of Bian'chi identities,
\begin{equation}
G_{\mu \nu} \equiv R_{\mu \nu}-\frac{1}{2} g_{\mu \nu}R=8 \pi G
T_{\mu \nu} -\Lambda g_{\mu \nu}
\end{equation}
The evolution equations in this case become,
\begin{eqnarray}
 &&H^2 = \frac{8\pi G}{3}\rho+\frac{\Lambda}{3}\\
&& \frac{\ddot{a}}{a}=-\frac{4\pi G}{3}\left(\rho +3p\right)
+\frac{\Lambda}{3} \label{aL}
\end{eqnarray}
In case the universe is dominated by $\Lambda$, it follows from the
continuity equation that $p_{\Lambda}=-\rho_{\Lambda}$ and
Eq.(\ref{aL}) tells us that a positive cosmological constant
contributes to acceleration.

Observations of complimentary nature reveal that,\\
$\bullet$ $\Omega_{tot} \simeq 1$~- CMB,\\
$\bullet$ $\Omega_m \simeq 0.3$~-Large scale structure and its dynamics,\\
$\bullet$ $\Omega_{DE}\simeq 0.7$~- high redshift Ia Supernove,\\
which is independently supported by data on baryon oscillations and
weak lensing.

Observations at present do not rule out the {\it phantom } dark
energy with $w<-1$ corresponding to super acceleration. In this case
the expanding solution takes the form,
\begin{eqnarray}
 a(t)=(t_s-t)^n ,~~(n=2/3(1+w))
\end{eqnarray}
where $t_s$ is an integration constant. It is easy to see that
phantom dominated universe will end itself in a singularity known as
{\it big rip} or {\it cosmic doomsday},
\begin{eqnarray}
&& H=\frac{n}{t_s-t} \, \\
&&R=6\left[\frac{\ddot{a}}{a}+\left(\frac
{\dot{a}}{a} \right)^2 \right]= 6\frac{n(n-1)+n^2}{(t_s-t)^{2}}
\end{eqnarray}
The big rip singularity is characterized by the divergence of $H$
and consequently of the curvature after a finite time in future. It
should be noted that when curvature becomes large, we should
incorporate the higher curvature corrections to Einstein-Hilbert
 action which can modify the structure of the singularity\cite{Abdalla}. Big rip
 can also be avoided in specific models of phantom energy\cite{param}.
 \subsection{Issues associated with $\Lambda$}
There are important theoretical issues related to cosmological
 constant.
Cosmological constant can be associated with vacuum fluctuations in
the quantum field theoretic context. Though the arguments are still
at the level of numerology but may have far reaching consequences.
Unlike the classical theory the cosmological constant in this scheme
is no longer a free parameter of the theory. Broadly the line of
thinking takes the following route.

The quantum effects in GR become important when the Einstein Hilbert
action becomes of the order of Planck's constant; this happens at
the Planck's length $Lp =    10^{-32}$ cm corresponding to Planck
energy which is of the order of $M^4_p \simeq    10^{72} GeV^4$. The
ground state energy dubbed zero point energy or vacuum energy of a
free quantum field is infinite. This contribution is related the
ordering ambiguity of fields in the classical Lagrangian and
disappears when normal ordering is adopted. Since this procedure of
throwing out the vacuum energy is {\it adhoc}, one might try to
cancel it by introducing the counter terms. The later, however
requires fine tuning and may be regarded as unsatisfactory. The
divergence is related to the modes of very small wave length. As we
are ignorant of physics around the Planck scale, we might be tempted
to introduce a cut off at $L_p$ and associate with this a
fundamental scale. Thus we arrive at an estimate of vacuum energy
$\rho_V \sim M_p^4$ (corresponding mass scale- $M_V \sim
\rho^{1/4}_V$) which is away by 120 orders of magnitudes from the
observed value of this quantity. The vacuum energy may not be felt
in the laboratory but plays important role in GR through its
contribution to the energy momentum tensor as $< T_{\mu\nu} >_0= -
\rho_V g_{\mu \nu}$ ($\rho_V=\Lambda/8\pi G, M_P^2=1/8\pi G$)and
appears on the right hand side of Einstein equations.

The problem of zero point energy is naturally resolved by invoking
supersymmetry which has many other remarkable features. In the
supersymmetric description, every bosonic degree of freedom has its
Fermi counter part which contributes zero point energy with opposite
sign compared to the bosonic degree of freedom thereby doing away
with the vacuum energy. It is in this sense the supersymmetric
theories do not admit a non-zero cosmological constant. However, we
know that we do not leave in supersymmetric vacuum state and hence
it should be broken. For a viable supersymmetric scenario, for
instance if it is to be relevant to hierarchy problem, the
suppersymmetry breaking scale should be around $M_{susy}\simeq 10^3$
GeV. We are still away from the observed value by many orders of
magnitudes. At present we do not know how Planck scale or SUSY
breaking scales is related to the observed vacuum scale.

 At present there is no satisfactory solution to
 cosmological constant problem. One might assume that there is some
 way to cancel the vacuum energy. One can then treat $\Lambda$ as a free
 parameter of classical gravity similar to Newton constant $G$.
 However, the small value of cosmological constant leads to several puzzles
 including the
  fine tuning and coincidence problems. The energy density in
  radiation at the Planck  scale is of the order of $10^{72} GeV^4$
  or $\rho_{\Lambda}/\rho_r \sim
  10^{-120}$
   Thus the vacuum energy density needs to be fine
  tuned at the level of one part in $10^{-120}$ around the Plank epoch,
  in order to match the current universe. Such an extreme fine
  tuning is absolutely unacceptable at theoretical grounds.
  Secondly, the energy density in cosmological constant is of the
  same order of matter energy density at the present epoch. The question what causes this
   {\it coincidence} has no satisfactory answer.

Efforts have recently been made to understand $\Lambda$ within the
frame work of string theory using flux compactification. String
theory predicts a very complicated landscape of  about $10^{500}$
de-Sitter vacua. Using Anthropic principal, we are led to believe
that we live in one of these vacua! It is easier to believe in God
than in these vacua!
\begin{figure}
\includegraphics[width=80mm]{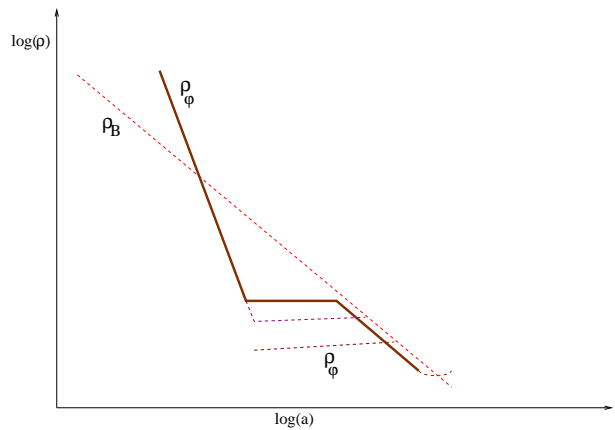}
\caption{Desired evolution of field energy density $\rho_{\phi}$
($\rho_B$ is the background energy density). The field energy
density in case of undershoot and overshoot joins the scaling
solution for different initial conditions. At late times, the scalar
field exits the scaling regime to become the dominant component. }
\label{shoot3}
\end{figure}
\begin{figure}
\includegraphics[width=80mm]{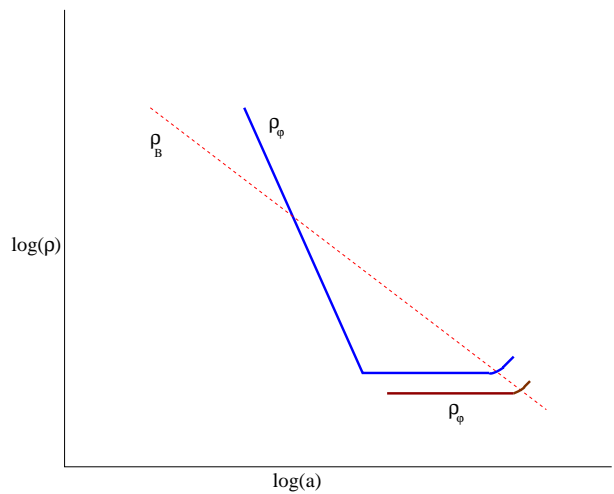}
\caption{Evolution of $\rho_{\phi}$ and $\rho_B$ in absence of the
scaling solution. The scalar field after its energy density
overshoots the background gets into locking regime and waits till
its energy density becomes comparable to $\rho_B$. It then begin to
evolve and over takes the background. similar picture holds in case
of the overshoot.} \label{shoot2}
\end{figure}

\section{Scalar field dynamics relevant to cosmology}
The fine tuning problem associate with cosmological constant led to
the investigation of cosmological dynamics of a variety of scalar
field systems such as quientessence, phantoms, tachyons and
K-essence. In past years, the underlying dynamics of these systems
has been studied in great detail. It is worthwhile to bring out the
broad features that makes a particular scalar field system viable to
cosmology. The scalar field model aiming to describe dark energy
should possess important properties allowing it to alleviate the
{\it fine tuning} and {\it coincidence} problems without interfering
with the thermal history of universe. The nucleosynthesis puts an
stringent constraint on any relativistic degree of freedom over and
above that of the standard model of particle physics. Thus a scalar
field has to satisfy several important constraints if it is to be
relevant to cosmology. Let us now spell out some of these features
in detail. In case the scalar field energy density $\rho_{\phi}$
dominates the background (radiation/matter) energy $\rho_B$, the
former should redshift faster than the later allowing radiation
domination to commence which in tern requires a steep potential. In
this case, the field energy density overshoots the background and
becomes subdominant to it. This leads to the locking regime for the
scalar field which unlocks the moment the $\rho_{\phi}$ is
comparable to $\rho_B$. The further course of evolution crucially
depends upon the form the potential assumes at late times. For the
non-interference with thermal history, we require that the scalar
field remains unimportant during radiation and matter dominated eras
and emerges out from the hiding at late times to account for late
time acceleration. To address the issues related to fine tuning, it
is important to investigate the cosmological scenarios in which the
energy density of the scalar field mimics the background energy
density. The cosmological solution which satisfy this condition are
known as {\it scaling solutions},
\begin{equation}
\frac{\rho_{\phi}}{\rho_B}=const.
\end{equation}
The steep exponential potential $V(\phi) \sim exp(\lambda \phi/M_P)$
with $\lambda^2>3(1+w_B)$ in the frame work of standard GR gives
rise to scaling solutions. Nucleosynthesis further constraints
$\lambda$. The introduction of a new relativistic degree of freedom
at a given temperature changes the Hubble rate which crucially
effects the neutron to proton for temperature of the order of one
MeV when weak interactions freeze out. This results into a bound on
$\lambda$, namely, $\Omega_{\phi}\equiv 3(1+w_B)/\lambda^2<0.13-0.2$
or $\lambda \gtrsim 4.5$. In this case, for generic initial
conditions, the field ultimately enters into the scaling regime, the
attractor of the dynamics and this allows to alleviate the fine
tuning problem to a considerable extent. The same holds for the case
of undershoot, see Fig.\ref{shoot3}.

Scaling solutions, however, are not accelerating as they mimic the
background (radiation/matter). One therefore needs some late time
feature in the potential. There are several ways of achieving this:
(1) The potential that mimics a steep exponential at early epochs
and reduces to power law type $V \sim \phi^{2 p}$ at late times
gives rise to accelerated expansion for $p<1/2$ as the average
equation of state $<w_{\phi}>=(p-1)/(p+1)<-1/3$ in this
case\cite{samioc1}. (ii) The steep inverse power law type of
potential which become shallow at large values of the field can
support late time acceleration and
 can mimic the background at early times\cite{samioc2}.

The solutions which exhibit the aforesaid features are referred to
as {\it tracker} solutions. For a viable cosmic evolution we need a
tracker like solution.

Recently, a variety of scalar field models such as tachyon and
phantom were investigated as candidates of dark energy. In case of
tachyon with equation of state parameter ranging from $-1$ to $0$,
there exists no scaling solution which could mimic the realistic
background (radiation/matter). Scaling solution which are possible
in this case are associated with negative equation of state and are
not of interest. In case of phantom scalar fields (scalar fields
with negative kinetic energy), there is no fixed point corresponding
to scaling solution. These scenarios suffer from the fine tuning
problem; dynamics in this case acquires dependence on initial
conditions\cite{scalar} (see Fig.\ref{shoot2}).

 The second approach to late tim acceleration is related to the modification of
 left hand side of Einstein equations or the geometry of space time. In the past
 few years, several schemes of large scale modifications
 have been actively investigated. In what follows, we shall briefly describe the modified theories of
gravity and their relevance to cosmology.
\section{Modified theories of gravity and late time acceleration}
In view of the above discussion, it is perfectly legitimate to
investigate the possibility of late time acceleration due to
modification of Einstein-Hilbert action. Some of these modifications
are inspired by fundamental theories of high energy physics where as
the others are based upon phenomenological considerations.
\subsection{String curvature corrections}
It is interesting to investigate the string curvature corrections to
Einstein gravity amongst which the Gauss-Bonnet correction enjoys
special status. These models, however, suffer from several problems.
Most of these models do not include tracker like solution and those
which do are heavily constrained by the thermal history of universe.
For instance, the Gauss-Bonnet gravity with dynamical dilaton might
cause transition from matter scaling regime to late time
acceleration allowing to alleviate the fine tuning and coincidence
problems. Let us consider the low energy effective action,
\begin{eqnarray}
S&=&\int{\rm{d}}^{4}x\sqrt{-g}\Big[\frac{1}{2\kappa^2}R
-(1/2)g^{\mu\nu}\partial_{\mu}\phi\:
\partial_{\nu}\phi-\nonumber \\
&-&V(\phi)-f(\phi)R_{GB}^2 \Big]+ S_{m} \label{Saction}
\end{eqnarray}
 where $ R_{GB}^2$ is the Gauss-Bonnet term,
\begin{eqnarray}
&& R_{GB}^2  \equiv R^2-4R_{\mu\nu}R^{\mu\nu}+
R_{\alpha\beta\mu\nu}R^{\alpha\beta\mu\nu}
\end{eqnarray}
The dilaton potential $V(\phi)$ and its coupling  to curvature
$f(\phi)$ are given by,
\begin{eqnarray}
&&V(\phi) \sim e^{(\alpha \phi)},~~f(\phi) \sim e^{-(\mu
 \phi)}
\end{eqnarray}
The cosmological dynamics of system (\ref{Saction}) in FRW
background was investigated in Ref.\cite{TS,KM06}. It was shown that
scaling solution ca be obtained in this case provided that $\mu=
\lambda$. In case $ \mu\ne \lambda$, we have the de-Sitter solution.
Hence, the string curvature corrections under consideration can give
rise to late time transition from matter scaling regime.
Unfortunately, it is difficult to reconcile this model with
nucleosynthesis\cite{TS,KM06}constraint.
\subsection{DGP model}
In DGP model, gravity behaves as four dimensional at small distances
but manifests its higher dimensional effects at large distances. The
modified Friedmann equations on the brane lead to late time
acceleration. The model has serious theoretical problems related to
ghost modes superluminal fluctuations. The combined observations on
background dynamics and large angle anisotropies reveal that the
model performs worse than $\Lambda CDM$\cite{roy}.
\subsection{Non-local cosmology} An
interesting proposal on  non-locally corrected gravity involving a
function of the inverse d'Almbertian of the Ricci scalar,
$f(\Box^{-1} R))$, was made in Refs.\cite{Wood} For a generic
function $f(\Box^{-1} R)\sim \exp(\alpha \Box^{-1} R)$, the model
can lead to de-Sitter solution at late times. The range of stability
of the solution is given by $1/3<\alpha<2/3$ corresponding to the
effective EoS parameter $w_{\rm eff}$ ranging as $ \infty <w_{\rm
eff}<-2/3$. For $1/3<\alpha<1/2$ and $1/2<\alpha<2/3$, the
underlying system is shown to exhibit phantom and non-phantom
behavior respectively; the de Sitter solution corresponds to
$\alpha=1/2$. For a wide range of initial conditions, the system
mimics dust like behavior before reaching the stable fixed point at
late times. The late time phantom phase is achieved without
involving negative kinetic energy fields. Unfortunately, the
solution becomes unstable in presence of the background
radiation/matter\cite{Wood}.
\subsection{f(R) theories of gravity}
On purely phenomenological grounds, one could seek a modification of
Einstein gravity by replacing the Ricci scalar by $f(R)$. The $f(R)$
gravity theories giving rise to cosmological constant in low
curvature regime are plagued with instabilities and on observational
grounds they are not distinguished from cosmological constant. The
recently introduced models of $f(R)$ gravity by Hu-Sawicki and
Starobinsky (referred as HSS models hereafter) with disappearing
cosmological constant\cite{HS,star} have given rise to new hopes for
a viable cosmological model within the framework of modified
gravity.
The action of  $f(R)$ gravity is given by\cite{review1},
\begin{equation}\label{action}
S = \int\left[\frac{f(R)}{16\pi G} + \mathcal{L}_m \right] \sqrt{-g}
\quad d^4 x,
\end{equation}
\noindent which leads to the following modified equations,
\begin{equation}\label{eq:freqn}
 f'R_{\mu\nu}-\nabla_{\mu\nu}{f'}+\left(\Box f' - \frac{1}{2}f\right)g_{\mu\nu}
=8\pi G T_{\mu\nu}.
\end{equation}
which are of  fourth order for a non-linear function f(R). Here
prime $(')$ denotes the derivatives with respect to $R$. The
modified Eq.(\ref{eq:freqn}) contains de-Sitter space time as a
vacuum solution provided that $ f(4\Lambda)=2\Lambda f'(4\Lambda)$.
Thus, the $f(R)$ theories of gravity may provide an alternative to
dark energy. The $f(R)$ gravity theories apart from a spin two
object necessarily contain a scalar degree of freedom. Taking trace
of Eq.(\ref{eq:freqn}) gives the evolution equation for the scalar
degree of freedom,
\begin{equation}\label{eq:frtrace}
 \Box f' = \frac{1}{3} \left ( 2f' - f' R \right ) + \frac{8\pi G}{3} T.
\end{equation}
\noindent It would be convenient to define scalar function $\phi$
as,
\begin{equation}
 \phi \equiv f' - 1,
\end{equation}
which is expressed through Ricci scalar once $f(R)$  is specified.
 We can write the trace equation (Eq.(\ref{eq:frtrace}))
 in the terms of $V$ and $T$ as
\begin{equation}
 \Box \phi = \frac{dV}{d\phi} +\frac{8\pi G}{3} T.
\end{equation}
\vskip 0.5cm \noindent The potential can be evaluated using the
following relation
\begin{equation}\label{Effpot}
 \frac{dV}{dR} = \frac{dV}{d\phi}\frac{d\phi}{dR}= \frac{1}{3}
\left ( 2 f - f' R \right )
 f''.
\end{equation}
The functional form of $f(R)$ should satisfy certain requirements
for the consistency of the modified theory of gravity. The stability
of
$f(R)$ theory would be ensured provided that,\\
$\bullet$ $f'(R)>$0 $-$~~~graviton~is~not~ghost,\\
$\bullet$ $f''(R)>0$$-$~~~scalaron~ is~ not~ tachyon.\\
The $f(R)$ models which satisfy the stability requirements can
broadly be classified into categories: (i) Models in which $f(R)$
diverges for $R\to R_0$ where $R_0$ finite or $f(R)$ is non
analytical function of the Ricci scalar. These models either can not
be distinguishable from $\Lambda CDM$ or are not viable
cosmologically. (ii) Models with $f(R)\to 0$ for $R\to 0$ and reduce
to cosmological constant in high curvature regime. These models
reduce to $\Lambda CDM$ in high redshift regime and give rise to
cosmological constant in regions of high density and differ from the
latter otherwise; in principal these models can be distinguished
from cosmological constant.

Models belonging to the second category were proposed by Hu-Sawicki
and Starobinsky  \cite{HS,star}. The functional form of $f(R)$ in
Starobinsky parametrization is given by,
\begin{equation}  \label{func}
f(R) = R + \lambda R_0 \left[\left(1+\frac{R^2}{R_0^2}\right)^{-n}
-1\right].
\end{equation}
 Here n and  $\lambda$ are positive. And $R_0$ is of the
order of presently observed cosmological constant, $\Lambda = 8\pi G
\rho_{vac}$. The properties of this model \cite{star} can be
summarized as follows:
\begin{enumerate}
\item  Stability conditions are satisfied as $f', f''>0$
\item   Flat space time is  an unstable solution of the model.
\item  For $|R| \gg R_0$, $f(R) = R - 2 \Lambda (\infty)$. The
high-curvature value of the effective cosmological constant is
$\Lambda(\infty) = \lambda R_0/2$.
\end{enumerate}
 In the Starobinsky model, the scalar field $\phi$ is given by
\begin{equation}\label{eq:phi}
 \phi(R) = -\frac{2n \lambda R}{R_0(1+\frac{R^2}{R_0^2})^{n+1}}.
\end{equation}
Notice that $R \to \infty$ for $\phi \to 0$. One can easily compute
$V(R)$ for a given value of $n$. For instance, in case of $n = 1$,
we have
\begin{eqnarray}
\frac{V}{R_0} &=& \frac{1}{24 \left ( 1 + y^2 \right )^4 } \left
\lbrace \left ( -8 - 40y^2 - 56y^4 -24y^6 \right ) \lambda + \left (
3y + 11y^3 + 21y^5 - 3y^7 \right ) \lambda^2
\right \rbrace \nonumber \\
&&- \frac{\lambda^2}{8}\tan^{-1}{y},
\end{eqnarray}
 where $y=R/R_0$.
 In the FRW background, the trace equation (Eq.
$(\ref{eq:frtrace})$) can be rewritten in the convenient form
\begin{equation}\label{eq:scalarfield}
 \ddot{\phi} + 3H\dot{\phi} + \frac{dV}{d\phi} =\frac{8\pi G}{3} \rho.
\end{equation}

\begin{figure}[htp]
\centering
\includegraphics[width=100mm]{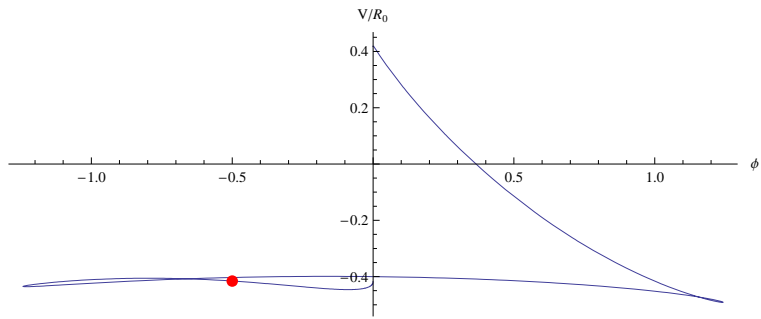}
\caption{Plot of effective potential for $n = 2$ and $\lambda =
1.2$. The red spot marks the initial condition for
evolution.}\label{fig:n2l1p2v}
\end{figure}
 The time-time component of the equation of motion
$(\ref{eq:freqn})$ gives the Hubble equation
\begin{equation}\label{eq:Friedmann}
 H^2 + \frac{d (\ln f')}{dt} H + \frac{1}{6}\frac{f - f'R}{f'}
= \frac{8\pi G}{3f'} \rho.
\end{equation}
It should be noted that the stability conditions ensure that the
effective gravitational constant $G_{eff}=G/f'$  appearing in
Eq.(\ref{eq:Friedmann}) is positive.
  The
simple picture of dynamics which appears here is the following:
above infrared modification scale $(R_{0})$, the expansion rate is
set by the matter density and once the local curvature falls below
$R_{0}$ the expansion rate gets effect of gravity modification.
 For pressure less dust, the effective potential has an extremum at,
\begin{equation}  \label{extm}
2 f - R f^{\prime }= 8\pi G \rho.
\end{equation}
For a viable late time cosmology, the field should be evolving near
the minimum of the effective potential. The finite time singularity
inherent in the class of models under consideration severely
constraints dynamics of the field.
\subsection*{The curvature singularity and fine tuning of parameters}
 The effective potential has minimum which depends upon $n$ and
$\lambda$. For generic values of the parameters, the minimum of the
potential is close to $\phi =0$ corresponding to infinitely large
curvature. Thus while the field is evolving towards minimum, it can
easily oscillate to a singular point. However, depending upon the
values of parameters, we can choose a finite range of initial
conditions for which scalar field $\phi$ can evolve to the minimum
of the potential without hitting the singularity.

\begin{figure}[htp]
\centering
\includegraphics[width=100mm]{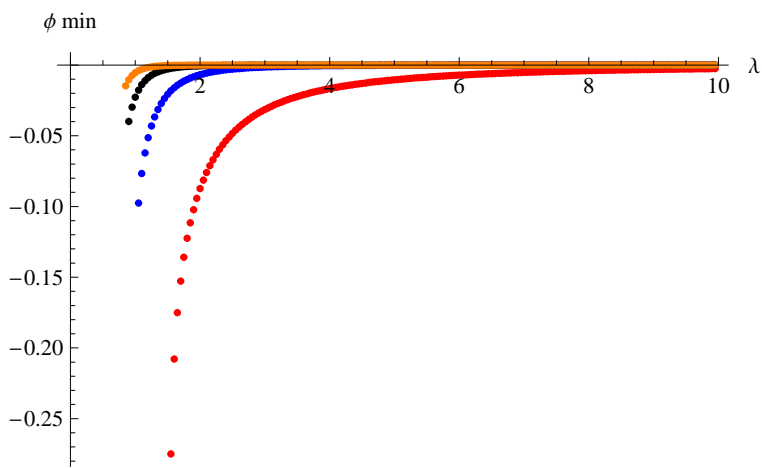}
\caption{Plot of $\phi_{min}$ versus $\lambda$ for different values
of $n$. With increase in $n$, $\phi_{min}$ approaches zero for
smaller values of $\lambda$. The curves from bottom to top
correspond to  $n = 1,2,3,4$ respectively.}\label{fig:phmin}
\end{figure}

 We find that the range of initial conditions allowed for the
evolution of $\phi$ to the minimum without hitting singularity
shrinks as the numerical values of parameters $n$ and $\lambda$
increase. This is related to the fact that for larger values of $n$
and $\lambda$, the minimum fast moves towards $\phi=0$, see figure
\ref{fig:phmin}. The numerical values should be accurately chosen to
avoid hitting the singularity.

\subsection*{Avoiding singularity with higher curvature corrections}
We know that in case of large curvature, the quantum effects become
important leading to higher curvature corrections. Keeping this in
mind, let us consider the modification of Starobinsky's model,
\begin{equation}
 f(R) = R + \frac{\alpha}{R_0} R^2 +R_0 \lambda \left \lbrack -1 + \frac{1}{(1+\frac{R^2}{R_0^2})^n} \right \rbrack,
\end{equation}
then $\phi$ becomes
\begin{equation}\label{eq:phi2}
 \phi(R) = \frac{R}{R_0}\left[2\alpha  -\frac{2n \lambda }{(1+\frac{R^2}{R_0^2})^{n+1}}\right].
\end{equation}

\begin{figure}[htp]
\centering
\includegraphics[width=60mm]{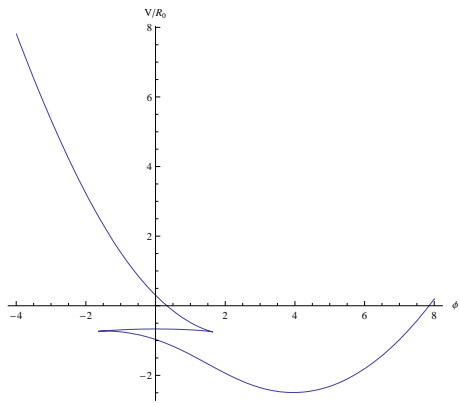}
\caption{Plot of the effective potential for $n =2, \lambda = 2$ and
$\alpha = 1/2$ in presence of $R^2$ correction. the minimum of the
effective potential in this case is located at $\phi_{min}=3.952
\,\,(R_{min} = 3.958)$.}\label{effpot}
\end{figure}

 When $|R|$ is large, the first term which comes from $\alpha R^2$
dominates. In this case, the curvature singularity $R = \pm \infty$
 corresponding to $\phi = \pm \infty$, see Fig.\ref{effpot}. Hence, in this modification, the
minimum of the effective potential is separated from the curvature
singularity by the infinite distance in the $\phi,V(\phi)$ plane.
 In case of $n = 2$, the expressions for $\phi$ and $V(\phi)$ are given
 by,
\begin{eqnarray}
 \phi(y) &=& 2\alpha y - \frac{4 \lambda y }{(1 + y^2)^3}, \\
\frac{V}{R_0} &=& -\frac{1}{480(1+y^2)^6} \left \lbrace \lambda^2 y
\left ( -105 - 595y^2 -2154 y^4 + 106 y^6
+ 595 y^8 + 105 y^{10}\right ) \right \rbrace \nonumber \\
&& -\frac{1}{3(1+y^2)^3} \left \lbrace 1 + 5y^2 + \alpha \left ( 3 +
8y^2 + 9y^4 + 4y^6 \right) \right \rbrace + \frac{1}{3} \alpha y^2 +
\frac{1}{32} \left (32 \alpha - 7\lambda \right ) \tan^{-1}(y).
\end{eqnarray}
 For $n = 2, \lambda = 2$ and $\alpha = 0.5$, we have a large range
of the initial condition for which the scalar field evolves to the
minimum of the potential. Though the introduction of $R^2$ term
formally allows to avoid the singularity but can not alleviate the
fine tuning problem as the minimum should be brought near to the
origin to respect the solar constraints. Last but not the least one
could go beyond the approximation (see Eq.(\ref{extm})) by iterating
the trace equation and computing the corrections to $R$ given by
equation (\ref{extm}). As pointed by Starobinsky\cite{star}, such a
correction might become large in the past. This may spoil the
thermal history and thus needs to be fine tuned. The aforesaid
discussion makes it clear that HSS models are indeed fine tuned and
hence very delicate.

In case of any large scale modification of gravity, one should worry
about the local gravity constraints. The $f(R)$ theories belong to
the class of scalar tensor theories corresponding the Brans-Dicke
parameter $\omega=0$ or the PPN parameter
$\gamma=(1+\omega)/(2+\omega)=1/2$ unlike GR where $\gamma=1$
consistent with observation ($|\gamma-|\lesssim 10^{-4}$). This
problem can be circumvented by invoking the so called chameleon
mechanism. In case, the scalar degree of freedom is coupled to
matter, the effective mass of the field depends upon the matter
density which can allow to avoid the conflict with solar physics
constraints.

\section{Summary}
We have briefly summarized here various approaches to understand the
late time acceleration of universe. In case of scalar field models
of dark energy, we emphasized the relevance of scaling solutions in
alleviating the fine tuning and the coincidence problems. We hope
that the future data would reveal the metamorphosis of dark energy.

Amongst the various alternatives to dark energy, the $f(R)$ gravity
models have received considerable attention in past years. There are
broadly two classes of $f(R)$ models, namely, those in which $f(R)$
diverges as $R\to R_0$ ($R_0$ is finite) or $f(R)$ is non-analytic
in $R$. And those with $f(R) \to 0$ as $R\to 0$, they reduce to
$\Lambda CDM$ in the limit of high redshift and give rise to
cosmological constant in high density regime. These models can evade
local gravity constraints with the help of the so called chameleon
mechanism and have potential capability of being distinguished from
$\Lambda CDM$\cite{david}.

Unfortunately, the $f(R)$  models with chameleon mechanism are
plagued with curvature singularity problem which may have important
implications for relativistic stars\cite{maed1}. The model could be
remedied with the inclusion of higher curvature
corrections\cite{TH}. At the onset, it seems that one needs to
invoke fine tunings to address the problem\cite{maed2}. The presence
of curvature singularity certainly throws a new challenge to $f(R)$
gravity models. In our opinion, the problem requires further
investigations. It would also be interesting to look for a realistic
scenario of quintessential inflation in the frame work of $f(R)$
gravity.
\section{Acknowledgements}
I thank A. Dev, D. Jain, S. Nojiri, S. Tsujikawa, S. Jhingan and I
Thongkool for useful discussions.

\end{document}